# The burden of HIV in a Public Hospital in Johannesburg, South Africa

Andrew Black,[1] Janie Kriel,[2] Michael Mitchley[3] and Brian G. Williams[1,4]

1. Wits Reproductive Health Institute, University of the Witwatersrand, Johannesburg, South Africa
2. Department of Medicine, University of the Witwatersrand, Johannesburg, South Africa
3. Computational and Applied Mathematics, University of the Witwatersrand, Johannesburg, South Africa
4. South African Centre for Epidemiological Modelling and Analysis (SACEMA), Stellenbosch, South Africa

Correspondence to Andrew Black: ablack@wrhi.ac.za

## Abstract

South Africa has the greatest number of people living with HIV in the world but the direct impact of this on the public health system has not been directly measured. Using data from the Chris Hani Baragwanath Hospital, the largest hospital in the Southern Hemisphere, collected between January 2006 and December 2009, we demonstrate directly the scale of the impact of HIV on mortality in health services in the pubic sector in South Africa. During the period under investigation 14,431 people died in the hospital's medical wards, an average of 11 deaths each day. Of those that died 64% of men and 82% of women were HIV positive. Between the ages of 30 and 40, 94% of men and 96% of women of those that died were HIV-positive. These data not only reflect the extraordinary mortality directly attributable to the epidemic of HIV but also the massive burden placed on the health services at a time when triple combination therapy was available and these HIV-related deaths could have been averted.

## Introduction

South Africa has, and continues to have, the greatest number of people infected with HIV in the world. Among adults aged 15 years or more, the annual incidence of HIV peaked at 2.5% in 1998, the prevalence of HIV reached 16% in 2006 and has not declined significantly since then, and the annual mortality peaked at 1.3% in 2007. The availability of ART in the public sector began in earnest in 2003 but by 2009 only 2.5% of adults were on ART, or about 16% of those infected with HIV. Since then the provision of ART has been scaled up rapidly and in 2015 about 10% of all adults are on ART or about 63% of all those living with HIV.[1]

For this study one of us (AB) collected detailed mortality data for everyone that died in the Chris Hani Baragwanath Hospital (CHBH) Medical wards between 2006 and 2009, during which time the incidence and the mortality reached their peak values and the rollout of ART in the public sector began. In 1996 triple combination therapy was already widely used in many countries and for many people infected with HIV, and/or those taking triple combination therapy, the excess mortality, was moderate and comparable with patients having other chronic conditions.[2] The data collected in this study therefore cover a critical period in the history of HIV in South Africa when the epidemic had reached its apogee but when effective treatment was widely available.

The impact of HIV not only on he lives and deaths of individual people but also the burden that it placed on the health services in the period from 2006 to 2009 are therefore of particular importance. In this paper we focus on the proportion of those that died who had been tested positive for HIV as a function of their age and gender. We will report the relationship between HIV status and different underlying causes of death in a later publication.

## Methods

In order to enable monitoring of deaths at CHBAH a data sheet was developed in 2006 to serve as a paper based mortality record. The data form was completed by a medical consultant at the time of signing the patients' death certificate. Cause of death was ascertained by the medical consultant by reviewing the patient file and the cause of death reported by the medical consultant may not be the same as that reported on the death certificate; because of the stigma surrounding HIV and HIV-related diseases HIV is not always recorded on the death certificate which is made available to patients families at CHBH. HIV status was reported as known positive, known negative, not known or clinically suspected because of AIDS-defining illness. If the HIV-status was not available in the patient records a search of the hospital laboratory records was done and the data sheet updated by an investigator (AB).

The CD4 cell count reported is the last available result on the hospital laboratory system, the vast majority having been taken during final admission. Antiretroviral therapy status was determined from patient records.

The data was captured into Microsoft Excel 2007 (Microsoft, Redmond, WA). Cause of death was coded using the Tenth International Classification of Diseases codes (ICD-10), by a person trained in ICD-10 coding. When queries arose a consultant (AB) assisted in assigning the code. Primary cause and possible secondary causes of death were coded but only the primary cause of death was used for this analysis.

The data for HIV-positive and HIV-negative men and women by age were fitted to skew-normal distributions so that $N(a)$, the number of people of age $a$, for each gender is given by

$$N(a) = \frac{\bar{N}}{\sigma\pi} e^{-\frac{(a-\mu)^2}{2\sigma^2}} \int_{-\infty}^{\alpha\left(\frac{a-\mu}{\sigma}\right)} e^{-\frac{t^2}{2}} dt \qquad 1$$

where $\bar{N}$ is the normalization parameter, $\mu$ is the location parameter, $\sigma$ is the scale parameter, and $\alpha$ determines the skewness of the distribution.

For those whose HIV status is known we fitted Equation 1 separately for each gender and for HIV-positive and HIV-negative people. For those whose HIV status was not known we fitted a curve of the form

$$N(a) = \pi N^+(a) + \nu N^-(a) \qquad 2$$

where we used the estimated location, scale and shape parameters for positive and negative men and women and varied the normalization parameters $\pi$ and $\nu$ to fit the estimated proportion of men and women, of unknown status, that are HIV-positive or HIV-negative.

## Results

Figure 1 shows the age distribution for HIV-positive men and women; the parameters from the fits are given in Table 1. As expected the peak age at death in HIV negative women (pink line) is 75 years while in HIV-negative men (light blue line) it is 65 years. However, the peak age at death in HIV-positive women is 33 years while in HIV-positive men it is 38 years. Not only does HIV massively increase mortality, with 82% of women and 64% of men being HIV-positive but the modal age of death in those that were HIV-positive is shifted down by 42 years in women and 27 years in men.

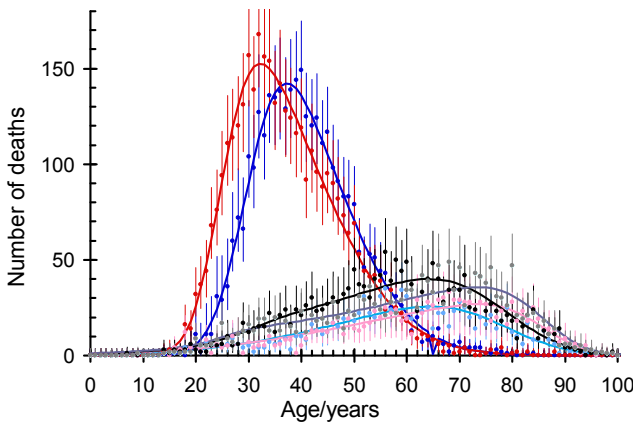

Figure 1. The number of men and women that died as a function of age. Red: HIV-positive women; Pink: HIV-negative women; Blue: HIV-positive men; Light blue: HIV-negative men; Black HIV-unknown men; Gray: HIV-unknown women.

The parameters for the fits are given in Table 1. For both men and women the distribution of those that are HIV-positive is shifted to younger ages and is much narrower. However, the age-distributions are skew to the left in those that are HIV-negative and skew to the right in those that are HIV-positive.

Using the fitted curves shown in Figure 1 and the parameters given in Table 1 we fitted the data for those whose HIV-status was not known and estimate the proportion of men and women who were infected with HIV among those whose status was unknown. This suggests that among those whose status was unknown only 6.7% of men and 6.8% of women were HIV-positive so that those of unknown status were overwhelmingly HIV-negative.

Table 1. Parameters from the fits in Figure 1 for men (M) and women (W) who are HIV-positive (+) or negative (–). $\bar{N}$ is the normalization factor, $\mu$ the location parameter, $\sigma$ the scale parameter and $\alpha$ the shape parameter as given in Equation 1.

|  | M– | M+ | W– | W+ |
|---|---|---|---|---|
| $\bar{N}$ | 4315 | 1415 | 4834 | 1571 |
| $\mu$ | 29.9 | 77.9 | 24.9 | 85.7 |
| $\sigma$ | 14.8 | 27.0 | 16.4 | 32.2 |
| $\alpha$ | 2.63 | −2.67 | 3.46 | −4.85 |

It is also instructive to plot the proportion of men and women who were HIV-positive as a function of age and this is shown in Figure 2. Between the ages of 21 and 43 years for women and 26 and 43 years for men, more than 90% of the deaths that occurred in Baragwanath were among those infected with HIV.

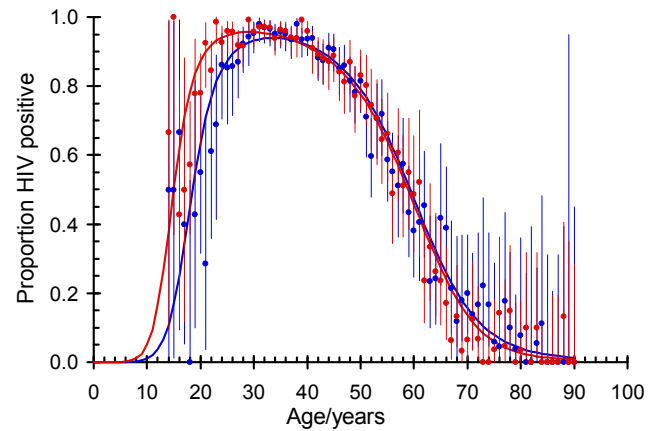

Figure 2. The proportion of men (blue dots and line) and women (red dots and line) who died and were HIV-positive as a function of age.

## Discussion

The epidemic of HIV has not only led to the deaths of several million people in South Africa but the mortality has been particularly severe among women aged 16 years to 60 years and among men aged 20 years to 60 years. Apart from the personal tragedies associated with each death, killing working aged adults will have had a very substantial impact on the economy of the country. Furthermore, in a country struggling to cope with a dysfunctional public health system the effect of HIV has been to increase the number of people dying in the major pubic hospital by four times. The recent recommendation by the World Health Organization to start people on ART as soon as possible after infection[3] should dramatically reduce not only AIDS related deaths but also the burden on the pubic health system.

## Conclusion

The tragedy is that South Africa, like many other countries, failed to provide triple combination therapy much earlier in the epidemic as this would have saved the lives of millions of young adults, minimized the economic impact of HIV, and very substantially reduced the burden on the health system. While it is true that in 2006 the cost of a course of ART was about US$730 per year it has now fallen to about US$100 per year[4] so that even maintaining 6 million people onto ART would only cost about US$600 million per year for the drugs or about 0.1% of the Gross Domestic Product.[5] Now that the WHO has embraced immediate treatment with ART for all those infected with HIV it is imperative that this is done to save lives, save money and make it possible to greatly reduce the burden on the health system thereby providing better health for all South Africans.

## Ethical approval

Ethical approval to analyse the operational data base was granted by The University of the Witwatersrand Human Research Ethics Committee, Clearance number M111103.

## Acknowledgements

This study received funding for the ICD 10 coding from the NICD/NHLS and was supported in part by funds from the United States Centers for Disease Control and Prevention (CDC), Atlanta, Georgia Preparedness and Response to Avian and Pandemic Influenza in South Africa (Cooperative Agreement Number: U51/IP000155-04). The contents are solely the responsibility of the authors and do not necessarily represent the official views of the CDC. The funders had no role in study design, implementation, manuscript writing or the decision to submit for publication.